\pdfoutput=1
\documentclass[conference]{IEEEtran}
\IEEEoverridecommandlockouts
\usepackage{cite}
\usepackage{amsmath,amssymb,amsfonts}
\usepackage{algorithmic}
\usepackage{graphicx}
\usepackage{textcomp}
\usepackage{xcolor}
\def\BibTeX{{\rm B\kern-.05em{\sc i\kern-.025em b}\kern-.08em
    T\kern-.1667em\lower.7ex\hbox{E}\kern-.125emX}}

\usepackage{eucal}
\usepackage[linesnumbered,ruled]{algorithm2e}
\SetAlCapNameFnt{\small}
\SetAlCapFnt{\small}
\usepackage{bbm}
\usepackage{float}
\usepackage{subcaption}
\usepackage{eurosym}
\usepackage{color}

\newcommand{\vu}{\mathbf{u}}
\newcommand{\vv}{\mathbf{v}}

\newcommand{\vx}{\mathbf{x}}
\newcommand{\vy}{\mathbf{y}}
\newcommand{\vz}{\mathbf{z}}

\newcommand{\valpha}{\boldsymbol{\alpha}}
\newcommand{\vbeta}{\boldsymbol{\beta}}
\newcommand{\vtheta}{\boldsymbol{\theta}}
\newcommand{\vdelta}{\boldsymbol{\delta}}
\newcommand{\vepsilon}{\boldsymbol{\epsilon}}
\newcommand{\vgamma}{\boldsymbol{\gamma}}
\newcommand{\vomega}{\boldsymbol{\omega}}

\newcommand{\mxh}{\mathbf{H}}

\newcommand{\mxx}{\mathbf{X}}

\newcommand{\mxsigma}{\boldsymbol{\Sigma}}

\begin{document}

\title{Probabilistic multivariate electricity price forecasting using implicit generative ensemble post-processing}

\IEEEpubid{\begin{minipage}{\textwidth}\ \\[12pt]
\copyright 2020 IEEE Personal use of this material is permitted. Permission from IEEE must be obtained for all other uses, in any current or future media, including reprinting/republishing this material for advertising or promotional purposes,creating new collective works, for resale or redistribution to servers or lists, or reuse of any copyrighted component of this work in other works.
\end{minipage}}

\author{
	\IEEEauthorblockN{Tim Janke and Florian Steinke}
	\IEEEauthorblockA{Energy Information Networks \& Systems\\
		Technische Universit{\"a}t Darmstadt\\
		\{tim.janke, florian.steinke\}@eins.tu-darmstadt.de}
}

\maketitle

\begin{abstract}
The reliable estimation of forecast uncertainties is crucial for risk-sensitive optimal decision making.
In this paper, we propose implicit generative ensemble post-processing, a novel framework for multivariate probabilistic electricity price forecasting.
We use a likelihood-free implicit generative model based on an ensemble of point forecasting models to generate multivariate electricity price scenarios with a coherent dependency structure as a representation of the joint predictive distribution.
Our ensemble post-processing method outperforms well-established model combination benchmarks.
This is demonstrated on a data set from the German day-ahead market.
As our method works on top of an ensemble of domain-specific expert models, it can readily be deployed to other forecasting tasks.
\end{abstract}


\section{Introduction}
In countries with liberalized electricity markets, prices play a central role for the efficient short-term coordination of supply and demand.
Operators of generation units, flexible loads, and storage facilities adapt their bidding and scheduling to anticipated market prices.
Hence, price forecasts are a crucial ingredient for flexible power systems.
Probabilistic forecasts extend classic point forecasts by also reporting the forecast uncertainty instead of only the expected value.
This allows an agent that bases his actions on the forecast to take risk-sensitive decisions by using tools from stochastic programming.
E.g. in \cite{Aasgard.2014} the operation of a multi-reservoir hydro power plant in the Nordic market under uncertain water inflow and market prices is modeled.
In \cite{Kumbartzky.2017} a combined heat and power plant which operates in the German market under uncertain day-ahead and balancing prices in considered.
\\
In these settings, electricity prices are a main source of uncertainty.
The optimization problem is usually solved using the sample average approximation of the objective \cite{Shapiro.2009} which requires samples from the multivariate joint predictive distribution.
Nevertheless, the probabilistic electricity price forecasting (EPF) literature is largely restricted to estimating marginal distributions over the single hourly prices \cite{Nowotarski.2018}.
However, sampling independently from the marginals would yield sub-optimal solutions since this ignores the dependency structure of the individual dimensions.
A common approach in other forecasting domains is to first estimate the marginal predictive distributions for the individual output dimensions and then generate samples from the joint distribution using copulas \cite{Schefzik.2018}.
\\
In this paper, we introduce implicit generative ensemble post-processing (IGEP) which bypasses these two steps by combining ideas from ensemble learning and implicit generative models (IGMs) \cite{Mohamed.2016}.
Our method works on top of an ensemble of deterministic expert models and allows to generate multivariate scenarios which represent the joint predictive distribution implicitly.
We use a multivariate linear model to transform samples from a set of univariate latent random variables and the deterministic mean forecasts to vectors of the target distribution.
The model is trained by minimizing the energy score (ES) \cite{Gneiting.2007a} using stochastic gradient descent.
Our model uses two types of stochastic latent variables, adaptive and non-adaptive.
The parameters of the adaptive latent variable distributions are set according to the ensemble dispersion and hence reflect the uncertainty of the mean prediction.
The parameters of the non-adaptive distributions have fixed parameters to reflect the second type of uncertainty, the unexplainable randomness.
\\
We demonstrate our approach on a publicly available data set of the German-Austrian day-ahead electricity market \cite{entsoe} where hourly prices are set via 24 simultaneous blind auctions under uniform pricing.
Our method outperforms various other approaches in terms of continuous ranked probability score (CRPS) and ES, including quantile regression averaging (QRA) \cite{Nowotarski.2015} and non-homogeneous Gaussian regression (NGR) \cite{Gneiting.2005,Jewson.2004} in combination with a Gaussian copula.
\\
The remainder of the paper is structured as follows.
We introduce main concepts regarding probabilistic forecasting, proper scoring rules, and IGMs in Section \ref{sec:RelatedWork}.
We then describe our approach theoretically in Section \ref{sec:IGEPs} and apply it to the problem of probabilistic EPF in Section \ref{sec:CaseStudy}.
We conclude in Section \ref{sec:Conclusion}.

\IEEEpubidadjcol
\section{Prior Work \& Concepts}\label{sec:RelatedWork}
\subsection{Multivariate Probabilistic Forecasting}

The literature on EPF has become extensive over the last two decades \cite{Weron.2014}.
In recent years the focus has shifted from point forecasts to probabilistic forecasts \cite{Nowotarski.2018}.
The goal of probabilistic forecasting is to correctly quantify prediction uncertainties.
Hence, probabilistic forecasts usually come in the form of a predictive conditional distribution over possible future quantities or events \cite{Gneiting.2014}.
\\
However, in order to use probabilistic forecasts for scenario-based stochastic optimization, the forecasts have to take the form of samples from the joint predictive distribution.
For univariate problems or if the output dimensions are independent this is straight forward.
In a multivariate setting, e.g. spatiotemporal modeling, sampling independently from the marginals would not lead to a coherent dependency structure.
In this case, the common approach is to first estimate a set of univariate, marginal predictive cumulative distribution functions (CDF) $F_d$ over $D$ individual output dimensions and then use copula functions to generate samples.
Sklar's theorem \cite{Sklar.1959} shows that every multivariate CDF $F$ with marginal distributions $F_1,...,F_D$ can be represented as
$F(y_1,...,y_D) = C(F_1(y_1),...,F_D(y_D))$ for $y_1,... ,y_D \in \mathbb{R}$.
The copula function $C:[0,1]^D \rightarrow [0,1]$ is a multivariate CDF with standard uniform marginals and allows to decouple the estimation of the marginal CDFs from determining the joint distribution.
A popular choice in many applications is the Gaussian copula.
Here, samples can be generated by first sampling a realization of $\vx \in \mathbb{R}^D$ from a $D$-dimensional standard normal distribution $\mathcal{N}(\mathbf{0},\mxsigma)$ with covariance matrix $\mxsigma$.
Then a predictive scenario $\hat{\vy}$ is obtained by setting $\hat{\vy} = [F^{-1}_1(\Phi(x_1)),...,F^{-1}_D(\Phi(x_D))]^T$, where $\Phi$ denotes the CDF of a univariate standard normal distribution and $F^{-1}_d$ is the inverse of the predictive CDF for dimension $d$.
\\
This approach is regularly applied in spatiotemporal problem settings like geostatistics, meteorological forecasting, and renewable energy forecasting \cite{Schefzik.2018}.
To our best knowledge only two EPF papers consider the problem.
Toubeau et al. \cite{Toubeau.2018} use empirical copulas to generate coherent scenarios of load, renewable generation, and prices for the Belgian market.
Chai et al. \cite{Chai.2018} employ a Gaussian copula approach for the Nordpool day-ahead market.
\subsection{Forecast Combination}
It is well known that combining forecasts from different models often improves accuracy.
The combination of probabilistic forecasting and ensemble methods is especially popular and well researched in the field of meteorological forecasting under the name of ensemble model output statistics (EMOS) \cite{Wilks.2018}.
In this domain, the task of model combination naturally arises from the need to statistically post-process the output of numerical weather forecasting ensembles to obtain a calibrated probabilistic forecasts.
However, ensemble post-processing methodologies and forecast averaging have also become popular in the probabilistic EPF literature, mainly as variants of QRA \cite{Nowotarski.2015}.
In QRA, forecasts from an ensemble of models are used as inputs for a quantile regression model to approximate the predictive distribution using a dense grid of quantiles.
However, there is to our best knowledge no EPF paper that systematically benchmarks other forecast combination schemes.

\subsection{Proper Scoring Rules}
In probabilistic forecasting we aim for predictive distributions that maximize sharpness subject to calibration.
Calibration refers to the consistency between the forecast and the observations, e.g. in expectation 20\% of the observed values should fall below the forecasted 0.2 quantiles, while sharpness refers to the concentration of the predictive distribution \cite{Gneiting.2007a}.
Scoring rules formalize the intuition of sharpness and calibration by assigning a numerical score to the predictive distribution depending on the realization of the random event or quantity of interest.
More formally, if we have access to samples $y \sim P$ from the true distribution $P$ and we issue a predictive distribution $Q$, let $S(Q,y)$ be our reward, where $S$ is a scoring rule.
A scoring rule is called strictly proper if its expected value is uniquely maximized for $P=Q$.
Thus, strictly proper scoring rules describe a principled framework for comparing probabilistic forecasts.
\\
The CRPS is a strictly proper scoring rule for real valued quantities that does not rely on a predefined likelihood.
It is defined as
\begin{align}\label{eq:CRPS}
	CRPS(F,y) = \int_{-\infty}^{\infty} (F(t) - \mathbbm{1}\{y \leq t \})^2 dt,
\end{align}
where $F$ is the CDF of the probabilistic forecast $Q$ \cite{Matheson.1976}.
Instead of computing the integral in (\ref{eq:CRPS}) we can also compute the CRPS by
\begin{align}\label{eq:CRPS_samples}
	CRPS(Q,y) =  \mathbb{E}_Q\left(|X-y|\right) - \frac{1}{2}\mathbb{E}_Q\left(|X-X'|\right),
\end{align}
where $X$ and $X'$ are independent samples of a random variable distributed according to $Q$ \cite{Gneiting.2007a}.
\\
Gneiting and Raftery \cite{Gneiting.2007a} introduce a generalization of the CRPS to the multivariate case.
The ES is defined as
\begin{align}\label{eq:ES}
	ES(Q,\vy) =  \mathbb{E}_Q \left(\Vert \mxx - \vy \Vert^\beta_2 \right) -\frac{1}{2} \mathbb{E}_Q \left( \Vert \mxx - \mxx' \Vert^\beta_2 \right)
\end{align}
and is strictly proper for $\beta \in (0,2)$.
Since the ES makes no assumption about the distributional form of $P$ and can be evaluated based on samples from the predictive distribution $Q$, it provides an attractive loss function for multivariate probabilistic forecasting tasks.

\subsection{Implicit Generative Models} 
Unlike prescribed generative models, which provide an explicit parametric description of the underlying probability distribution, IGMs are likelihood-free models that only define a stochastic procedure to generate samples \cite{Mohamed.2016}.
Since we are primarily interested in sampling from the predictive distribution, IGMs form an attractive class of models for multivariate probabilistic forecasting.
\\
A prominent example of IGMs are GANs \cite{Goodfellow.2014} which are based on deep neural networks and replace the likelihood function with a classification model called the discriminator.
The only input for the generator model are vectors of unconditional random noise from a set of simple, univariate latent variable distributions, e.g. independent uniform distributions.
GANs have already been applied to probabilistic wind power forecasting in \cite{Chen.2018} where they are used to generate unconditional error scenarios for multi-step ahead forecasting.
While GANs are potentially very powerful models and achieve impressive results on a variety of tasks, they are known for a complex and unstable training procedure and require large training data sets \cite{Goodfellow.2016}.
MMD-GANs \cite{Dziugaite.2015,Li.2015} replace the discriminator model with the maximum mean discrepancy two-sample test (MMD) \cite{Gretton.2012} which significantly simplifies the training procedure.
Interestingly, the ES is a special case of the MMD \cite{Sejdinovic.2013}.

\section{Implicit Generative Ensemble Postprocessing}\label{sec:IGEPs}

\subsection{Problem Setting}
Consider a data set comprised of $N$ examples $\{(\vx,\vy)_n\}_{n=1}^N$, where $\vy \in \mathbb{R}^{D}$ is the output and $\vx=[\vx_1,...,\vx_M]$ are previously determined point predictions for $\vy$ from a set of $M$ different models, with $\vx_m \in \mathbb{R}^D$ and $\vx \in \mathbb{R}^{D \times M}$.
We denote the mean prediction of the ensemble members by $\bar{\vx} = \frac{1}{M}\sum_m \vx_m$, i.e. $\bar{\vx} \in \mathbb{R}^{D}$.
Our goal is to generate a set of $S$ scenarios $\{\hat{\vy}^s\}_{s=1}^S$ that represent the predictive joint distribution $p(\vy|\vx)$.
In order to do this, we parametrize a generator function $G_{\vtheta}(\vx,\vz^s)$ with model parameters $\vtheta$, where $\vz^s \in \mathbb{R}^K$ denotes a sample from $K$ independent univariate latent distributions.

\subsection{Latent Variable Distributions}
The latent variables in IGMs are usually treated as independent random noise and serve as an external source of randomness to generate samples from the unconditional implicit distribution.
We propose to give meaning to some of the latent distributions by adapting their parameters depending on the ensemble predictions.
We construct $D$ uniform adaptive latent distributions $\mathcal{U}_d(-\delta_d,+\delta_d)$, where 
\begin{align}
\delta_d = (\max(\vx_d) - \min(\vx_d))/2.
\end{align}
and $x_d=[x_{d,1},...,x_{d,M}]$.
This will reduce the variance of the samples for the dimension $d$ if the ensemble predictions are similar and increase the variance in case the ensemble is more dispersed.
We denote a sample from the adaptive latent distributions by $\vu^s$.
\\
The uncertainty in the final prediction is only partly due to the uncertainty in the mean predictions.
It is also caused by effects we are not able to model and hence treat as random noise.
We therefore specify $J$ additional latent distributions $\mathcal{U}_j(-1,+1)$ that are independent of the point predictions, i.e. they have a constant variance.
We denote a sample from the non-adaptive latent distributions by $\vv^s$.
To construct a sample from the latent space, we draw from all latent distributions and concatenate the values to the vector $\vz^s=\begin{bmatrix} \vu^s \\ \vv^s \end{bmatrix}$.

\subsection{Generator Function}
\begin{figure}
	\centering
	\includegraphics[width=3.55in]{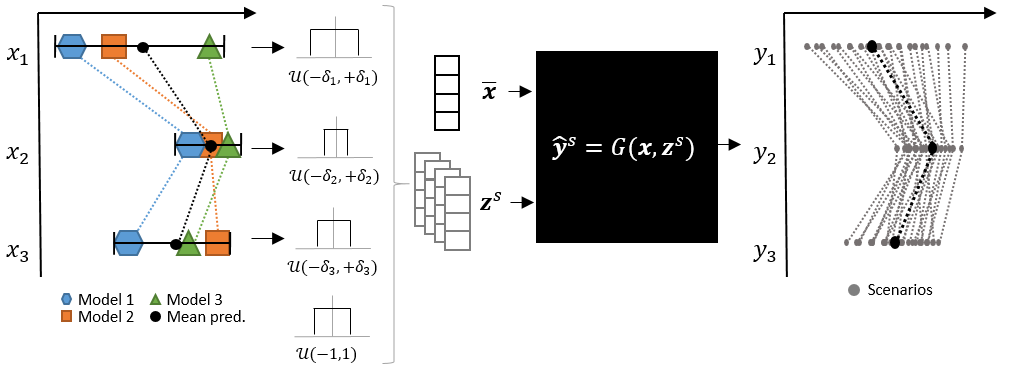}
	\caption{For a single example $(\vx,\vy)$  a set of $S$ scenarios is generated by first adapting the parameters of $D$ latent distributions according to the ensemble dispersion. Then a set of samples $\{\vz^s\}_{s=1}^S$ from all $K$ latent distributions is drawn and passed through the model to generate a set of scenarios $\{\hat{\vy}^s\}_{s=1}^S$.} \label{Model}
	\label{fig:Model}
\end{figure}

To be able to model a coherent dependency structure, the architecture of the generator function must ensure that all elements in $\hat{\vy}^s$ can depend on the whole latent random vector $\vz^s$.
In general, $G$ could be any differentiable function, e.g. a deep neural network.
However, as training data in a lot of forecasting tasks is usually relatively small, we propose a linear model of the form
\begin{align}\label{eq:model_def}
	\hat{\vy}^s = G_{\vtheta}(\vx,\vz^s) = \valpha + \vbeta \odot \bar{\vx} + \vgamma \vu^s + \vomega \vv^s,
\end{align}
with $\valpha \in \mathbb{R}^D, \vbeta \in \mathbb{R}^D$, $\vgamma \in \mathbb{R}^{D \times D}$, and $\vomega \in \mathbb{R}^{D \times J}$.
Thus, the model has $2D + DD + DJ$ parameters.
The first two terms allow for bias correction of the deterministic ensemble mean. The third term models the dependence on the adaptive latent variables while the fourth term models the dependence on the non-adaptive latent variables.
Hence, for output dimension $d$, the prediction $\hat{y}_d$ is a linear combination of the respective ensemble mean $\bar{x}_d$ and all elements in $\vu^s$ and $\vv^s$,
\begin{align}\label{eq:model_def_2}
	\begin{split}
	\hat{y}_d^s = \alpha_{d} + \beta_{d} \bar{x}_d &+ \gamma_{d,1} u^s_1+ ... + \gamma_{d,d} u^s_D 
	\\ &+ \omega_{d,1} v^s_1+ ... + \omega_{d,J} v^s_J.
	\end{split}
\end{align}
Note that by construction $\mathbb{E}(\vu) = \mathbf{0}$ and $\mathbb{E}(\vv) = \mathbf{0}$ and therefore $\mathbb{E}(\hat{\vy}) = \valpha + \vbeta \odot \bar{\vx}$ as can be seen from (\ref{eq:model_def}) and (\ref{eq:model_def_2}), i.e. in expectation the model predicts the bias corrected ensemble mean.
\\
Fig. \ref{fig:Model} shows the process of generating a set of scenarios for a single example $(\vx,\vy)$. First, the parameters of $D$ latent distributions are adapted according to the ensemble dispersion $\vdelta = [ \delta_1,..., \delta_D]^T$. We then draw $S$ samples $\{\vz^s\}_{s=1}^S$ from the univariate latent distributions.
Together with the mean prediction $\bar{\vx}$, each of these samples is then mapped to the multivariate output space via $G_{\vtheta}(\bar{\vx},\vz^s)$.
We thereby obtain a set of $S$ scenarios $\{\hat{\vy}^s\}_{s=1}^S$ which represent the predictive distribution.

\subsection{Training}
We train $G$ to minimize the expected ES given by
\begin{align}\label{eq:LossFcn}
	\begin{split}
		\mathcal{L} =& \frac{1}{N} \sum_n \bigg [ \frac{1}{S} \sum_s \Vert \vy_n- G_{\vtheta}(\vx_n,\vz_n^s) \Vert^{\beta}_2 \\
		& - \frac{1}{2S(S-1)} \sum_s \sum_{s'\neq s} \Vert G_{\vtheta}(\vx_n,\vz_n^s) - G_{\vtheta}(\vx_n,\vz_n^{s'}) \Vert^{\beta}_2 \\
		& + \lambda \Vert \vtheta \Vert_F^2 \bigg]
	\end{split}
\end{align}
with $\beta=1$ and $\Vert . \Vert_F$ is the Frobenius norm.
Intuitively, this objective function represents two orthogonal goals.
The first term in (\ref{eq:LossFcn}) decreases if the generated scenarios are close to the true value while the second term increases when the distance between the scenarios is large and hence rewards scenarios that are diverse.
The last term is a regularization term that is controlled by the parameter $\lambda$.
A sensible initialization for the model parameters is $\valpha=\mathbf{0}, \vbeta=\mathbf{1}$, $\vgamma = \mathbf{I}$, and $\vomega = \mathbf{0}$.
This corresponds to a model that in expectation predicts the ensemble mean and the uncertainty of dimension $d$ only depends on the ensemble spread for $d$.
During training, only a small number of scenarios $S_{train}$ is generated per training example.
We train the model using stochastic gradient descent with batch size $N_b$ and use automatic differentiation to compute the gradients.
Algorithm \ref{alg:Training} formalizes the training procedure.
\begin{algorithm}
	\small
	\SetKwInOut{Input}{Input}
	\SetKwInOut{Output}{Output}
	\Input{data {$\{\vx,\vy\}_{n=1}^N$, initial parameters $\vtheta_0$}, no. of ind. latent variables $J$, no. of samples during training $S_{train}$, batch size $N_b$, learning rate $\eta$}
	\Output{Model parameters $\vtheta^*$}
	\For{$N_{epochs}$}{
		Get batch $\{(\vx,\vy)_{n_b}\}_{n_b=1}^{N_b}$ \\
		\For{$n_b=1,...,N_b$}{
			Set parameters of adaptive latent distributions according to $\vdelta_{n_b}$ \\
			\For{$s=1,...,S_{train}$}{
				Generate a latent sample $\vz^s$ \\
				Generate a scenario $\hat{\vy}^s$ via $G_{\vtheta}(\vz^s,\vx_{n_b})$
			}
		}
		Compute gradient $\nabla_{\vtheta} \mathcal{L}$ over batch\\
		Update learning rate $\eta$\\
		Update model parameters $\vtheta \leftarrow \vtheta -  \eta \nabla_{\vtheta} \mathcal{L}$	
	}
	\caption{Training algorithm}
	\label{alg:Training}
\end{algorithm}

\section{Case Study: Probabilistic Electricity Price Forecasting}\label{sec:CaseStudy}

\subsection{Forecasting Study}
In the following, we demonstrate how to apply our approach to the task of probabilistic electricity price forecasting using a publicly available data set of the German-Austrian day-ahead market from January 2015 to December 2017 \cite{entsoe}.
As in most European countries, this market is operated as a daily blind auction under uniform pricing for 24 one-hour-blocks of electrical energy that is to be consumed or delivered during the respective hour of the following day.
The data set contains the prices $y_{d,h}$ for the hours $h \in \{1,...,24\}$ on the days $d \in \{1,...,1096\}$ along with the forecasted load $Load_{d,h}$, wind power generation $Wind_{d,h}$, and solar power generation $PV_{d,h}$.
The forecasting task is to issue a probabilistic forecast for the price vector $\vy_{d} \in \mathbb{R}^{24}$ in form of a set of samples $\{\hat{\vy}_{d}^s\}_{s=1}^S$.
To apply our model, we first have to construct a set of point forecasting models that form the ensemble.
The out-of-sample forecasts of the ensemble are then used as inputs for the probabilistic models.
We use the data of 2015 as initial training set for the ensemble models.
The ensemble forecasts for 2016 then form the training set for the probabilistic models.
We use the full year of 2017 as test set.
For all models, we apply a rolling window scheme, i.e. after forecasting the values for the next day, the training set is shifted by one day and all models are reestimated.

\subsection{Point Forecasting Ensemble}
We train a set of $M=5$ expert point forecasting models and model the prices as a function of the residual load $RL_{d,h} = Load_{d,h} - Wind_{d,h} - PV_{d,h}$, i.e. the share of the demand that is not covered by generation from wind and solar.
Point forecasts are denoted by $x_{d,h}$.

\paragraph{ARX-M}The first model is an ARX-type linear regression model given by $\tilde{x}_{d,h} = w_0 + w_1RL_{d,h} + w_2\tilde{y}_{d-1,h} + w_3\tilde{y}_{d-2,h} + w_4\tilde{y}_{d-7,h} +  w_5RL_{d-1,h} + w_6RL_{d-2,h} + w_7RL_{d-7,h}$.
This model works on $asinh$-transformed prices \cite{Uniejewski.2017} denoted by $\tilde{y}$ to account for the non-linear effect of the residual load.
We fit one model per hour of the day, i.e. we fit 24 separate models.

\paragraph{ARX-U} The second model is of similar type but we only fit a single model for all hours of the day.
The model is given by $\tilde{x}_{d,h} = w_1RL_{d,h} + w_2\tilde{y}_{d-1,h} + w_3\tilde{y}_{d-2,h} + w_4\tilde{y}_{d-7,h} +  w_5RL_{d-1,h} + w_6RL_{d-2,h} + w_7RL_{d-7,h} + \sum_{i=1}^{24} w_{7+i}H_{d,h,i} $, where  $\mxh = [H_1,...,H_{24}]^T$ is a one hot encoded vector of hour dummies.

\paragraph{Poly-LR} This model is a polynomial linear regression model given by $x_{d,h} = w_1RL_{d,h} + w_2RL_{d,h}^2 + w_3RL_{d,h}^3 + \sum_{i=1}^{24} w_{3+i}H_{d,h,i}$.
This model does not use any lagged predictors.
To still account for autocorrelation, we estimate the model parameters using weighted least squares, i.e. we weight the $i$th training sample by $\kappa_i = \exp(-0.01(d-d_i)^2)$.

\paragraph{LW-LR} This model is a locally weighted linear regression model given by $x_{d,h} = w_0+w_1RL_{d,h}$ with weights $\kappa_i = \exp(-0.01(d-d_i)^2-10(RL_{d,h}-RL_{i})^2)$.

\paragraph{GB} The last model uses gradient boosted decision trees and models the price as a function of the residual load and the hour dummies $x_{d,h} = f(RL_{d,h},\mxh_{d,h})$.
It is also estimated using a weighted training set with weights $\kappa_i = \exp(0.01(d-d_i)^2)$.
The model was estimated using Scikit-learn 0.20.1 with all hyperparameters kept at their default values.
\\
We report the mean absolute error (MAE) and the root mean squared error (RMSE) of the models' forecasts and the simple average of all forecasts (AVG) $\bar{x}_{d,h}=\frac{1}{M}\sum_{m}x_{d,h}^m$ for the entire out of sample period in Table \ref{tab:results_point}.
Taking the average of all forecasts reduces the MAE by 5\% and RMSE by 2\% in comparison to the best performing model.
\begin{table}[h]
	\centering
	\caption{Out-of-sample (2016+2017) MAE and RMSE values in $EUR/MWh$ for the point forecasting models}
	\label{tab:results_point}
	\begin{tabular}{c c c c c c c}
		\hline
		& ARX-U	&ARX-M 	& Poly-LR & LW-LR  & GB    & AVG  \\ 
		\hline
		MAE \quad  	& 3.65 & 3.64   & 3.46	  & 3.64  & 3.58   & \textbf{3.29} \\
		RMSE \quad  & 6.08 & 6.08	& 5.53 	  & 5.86   & 6.17  & \textbf{5.42} \\
		\hline
	\end{tabular} 
\end{table}

\subsection{Model Benchmarking}
We set up an IGEP model as described in Section \ref{sec:IGEPs} with hyperparameters $N_b=3, S_{train}=25, \lambda=0,$and $J=10$, i.e. the model has 864 parameters.
Before training we standardize all inputs and outputs using the mean and standard deviation of the prices in the training set.
We train the model for 100 epochs using the Adam optimizer \cite{adam} at default values in Keras 2.2.4 \cite{Keras}. 
Training the model takes about one minute on a standard laptop with an Intel i7-7500U CPU.
We generate 1000 scenarios for each of the 365 days in the test set and evaluate our model by computing the average ES and CRPS.
The whole training and test procedure was repeated 10 times for all models to average out random effects during training and evaluation.
We compare the results of our model against the five approaches described below.

\paragraph{Raw Ensemble}
The simplest approach is to treat the $M$ ensemble members as $M$ samples from the predictive distribution, i.e. we set $\{\hat{\vy}_d^s\}_{s=1}^S = \{\vx_d^m\}_{m=1}^M$.

\paragraph{Multivariate Gaussian Errors (MGE)}
We can easily generate scenarios $\hat{\vy}_d^s = \bar{\vx}_d + \vepsilon^s,$ where $\vepsilon \sim \mathcal{N}(\mathbf{0},\mxsigma_{{\vepsilon}})$ and $\mxsigma_{\vepsilon}$ is the covariance matrix of the residuals of the ensemble mean predictions $\bar{\vx}$ from the training set.
This method can generate samples with a coherent dependency structure if the homoscedastic multivariate Gaussian error assumption is appropriate.

\paragraph{IGEP with independent latent variables (IGEP\textsubscript{ind})}
Here we use the IGEP model with the same hyperparameters but independent latent variables, i.e. we set $\delta_d=2$ for all examples.
Thus, the variance of the latent distributions is not adapted to the ensemble spread and the model does not account for uncertainty in the ensemble mean prediction.
However, the model should be able to generate realistic price scenarios if the model's capacity is sufficient.

\paragraph{Quantile Regression Averaging \& Gaussian copula (QRA+C)}
We use quantile regression averaging (QRA) \cite{Nowotarski.2015} to approximate the marginal predictive distribution for each price by a dense grid of quantiles $\tau \in \{0.01,...,0.99\}$.
QRA applies a linear quantile regression \cite{Koenker.1978} model  $\hat{Q}_{d,h}(\tau) = \beta_0 + \beta_1 x_{d,h}^1 + ... + \beta_M x_{d,h}^M$ to an ensemble of $M$ point forecasts, where $\hat{Q}(\tau)$ is the predicted value for quantile $\tau$.
We then use the Gaussian copula approach described in \cite{Pinson.2009} with the standardize error covariance matrix $\tilde{\mxsigma}_{\vepsilon}$ to generate scenarios based on the approximated univariate marginal distributions.
QRA is a well established and strong benchmark for probabilistic electricity price forecasting \cite{Nowotarski.2018}.
Also note that the pinball loss function minimized in QRA is an approximation of the CRPS.

\paragraph{Nonhomogeneous Gaussian Regression \& Gaussian copula (NGR+C)}
NGR \cite{Gneiting.2005,Jewson.2004} is an approach originally developed for post-processing numerical weather forecast ensembles.
The marginal probabilistic forecast takes the form of a univariate Gaussian distribution $\mathcal{N}(\hat{\mu}_{d,h},\hat{\sigma}_{d,h}^2)$.
The mean and variance parameters are estimated using two different models.
We use the model $\hat{\mu}_{d,h}=\beta_0 + \beta_1\bar{x}_{d,h}$ to estimate the mean and the model $\hat{\sigma}_{d,h} = \gamma_0 + \gamma_1 s_{d,h}$ to estimate the variance, where $s_{d,h}$ is the standard deviation of the ensemble predictions.
We estimate the NGR model parameters using maximum likelihood (ML) as well as minimum CRPS estimation using the R $crch$ package \cite{Messner.2016}.
We then again use the Gaussian copula to generate multivariate samples.

\begin{figure}
	\begin{subfigure} {0.99\textwidth}
		\includegraphics[width=3.5in]{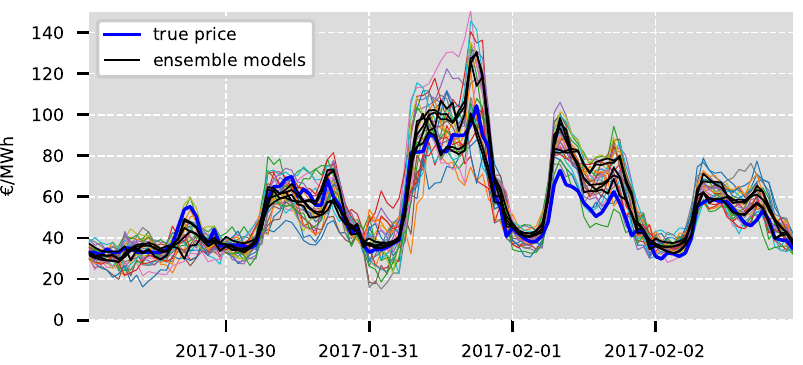}
		\label{subfig:S1}
	\end{subfigure}
	\begin{subfigure} {0.99\textwidth}
		\includegraphics[width=3.5in]{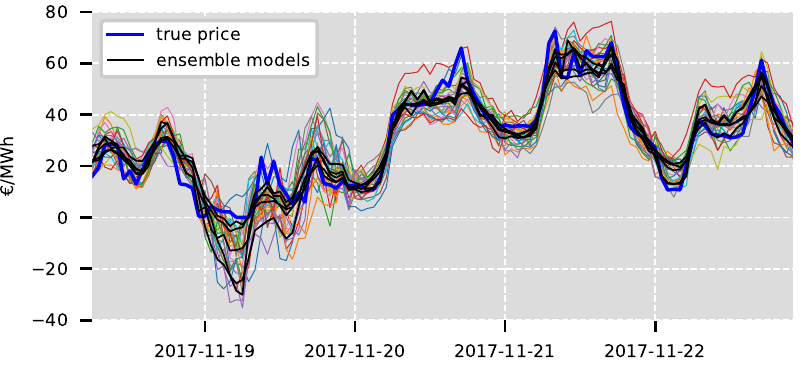}
		\label{subfig:S2}
	\end{subfigure}
	\caption{Generated scenarios (colored), true price (blue), and ensemble point predictions (black) for two different market situations. The model generates a more diverse set of scenarios if the ensemble dispersion is larger.}
	\label{fig:Scenarios}
\end{figure}

\subsection{Results}
\begin{table*}[t]
	\centering
	\caption{Mean test set ES, CRPS, and RMSE values in $EUR/MWh$ and standard deviations from 10 independent training and test runs.}
	\label{tab:results}
	\begin{tabular}{c c c c c c c c}
		\hline
		& IGEP 	  					   & Raw Ens. 		&MGE             & IGEP\textsubscript{ind}& QRA+C 		   & NGR\textsubscript{ML}+C & NGR\textsubscript{CRPS}+C\\ 
		\hline
		ES 	  & \textbf{16.294}$\pm 0.008$ &18.591 $\pm0$	&17.261 $\pm0.016$  & 16.995$\pm0.007$ & 16.806$\pm0.008$ & 16.427$\pm0.009$ 	& 16.410$\pm0.005$\\
		CRPS  & \textbf{2.693}$\pm0.002$   &3.052 $\pm0$ 	&2.866  $\pm0.003$  & 2.799$\pm0.002$  & 2.783$\pm0.002$ 	   & 2.716$\pm0.002$ 		& 2.704$\pm0.001$\\
		RMSE  & 6.014$\pm0.002$ & 5.994$\pm0$ & 5.994$\pm0$ & 6.000 $\pm0.002$ & \textbf{5.956}$\pm0$ & 6.022$\pm0$ & 6.030$\pm0$\\
		\hline
	\end{tabular} 
\end{table*}

Fig. \ref{fig:Scenarios} shows scenario forecasts generated by the IGEP model for two different market situations together with the true price and the ensemble predictions.
As can be seen, the model generates samples that resemble the real price vectors.
Furthermore, it generates more diverse scenarios at times where the ensemble spread is increased.
\\
Table \ref{tab:results} shows the test set results for the average ES and CRPS values as well as the RMSE for the predictive mean.
Recall that the CRPS does not consider the joint dependency structure like the ES but only accounts for the marginal predictive distributions.
Hence, the CRPS values of the QRA and NGR models are only dependent on the models for the marginals while the ES also depends on the employed copula.
All tested models improve over the ES and CRPS values of the raw ensemble with the MGE approach showing the smallest improvement.
This indicates that the assumption of homoscedastic Gaussian errors is not appropriate.
The IGEP model performs best in terms ES and CRPS.
It shows a 4 \% improvement over the IGEP\textsubscript{ind} model in terms of ES.
This improved performance must result from the adaptive latent variable distributions as the models are otherwise identical.
Hence, the IGEP model is capable of meaningfully transforming the random samples from the univariate adaptive latent distributions to samples of the multivariate predictive distribution.
The second best model is the NGR model with minimum CRPS estimation which slightly improves over the NGR model with maximum likelihood estimation.
Interestingly, QRA, which is quite popular in the EPF literature, shows the worst performance of the more advanced methods.
While the differences in the scores seem small, note that the differences must largely result from a better assessment of forecasting uncertainty since the RMSE values for all models are very similar.


\section{Conclusion}\label{sec:Conclusion}
In this paper we proposed IGEP, a novel forecast combination method for scenario-based multivariate probabilistic forecasting that combines IGMs with an ensemble of point prediction models.
Using a given set of point forecasts as inputs, our method allows to generate multivariate samples from the joint predictive distribution without making parametric assumptions about the underlying probability distribution.
We demonstrated our approach for the task of probabilistic electricity price forecasting.
Our method outperformed two well established benchmarks, QRA and NHR in combination with a Gaussian copula, in terms of ES and CRPS.
Since our method works on top of an ensemble of domain specific expert models, it can be applied in a wide variety of forecasting tasks.
\\
There are several avenues for future work.
We did not systematically investigate the effect of changing the model's hyperparameters, e.g. the number and type of latent variable distributions.
Introducing a non-linear model structure is also worth investigating.
Furthermore, we plan to test and compare the proposed framework on more data sets.
In general, we see a lot of potential in combining IGMs and expert models for probabilistic forecasting, especially if one wants to avoid making parametric assumptions about the predictive distribution.


\bibliographystyle{IEEEtran}
\bibliography{References}

\end{document}